\documentclass[twocolumn,showpacs,preprintnumbers,amsmath,amssymb,nofootinbib]{revtex4}

\usepackage[dvips]{graphicx}
\usepackage{subfigure}
\usepackage{rotating}
\DeclareGraphicsExtensions{.eps}

\def\eq{\begin{equation}}
\def\qe{\end{equation}}
\def\eqa{\begin{eqnarray}}
\def\qea{\end{eqnarray}}

\newcommand{\newc}{\newcommand}

\newc{\softsusy}{\texttt{SOFTSUSY}}

\newc{\ifb}{\textrm{fb}^{-1}}

\newc{\MO}{M_0}
\newc{\Mhalf}{M_{1/2}}
\newc{\AO}{A_0}
\newc{\tanb}{\textrm{tan}\beta}
\newc{\sgnmu}{\textrm{sgn}{\mu}}

\newc{\bbdecay}{0\nu\beta\beta}
\newc{\betadecay}{0\nu\beta\beta}
\newc{\bhalflife}{T^{0\nu\beta\beta}_{1/2}(\Ge)}
\newc{\mee}{m_{\beta\beta}}
\newc{\Ge}{^{76}\textrm{Ge}}
\newc{\bbbar}{B^0_d\textrm{-}\bar{B}^0_d}
\newc{\kkbar}{K^0\textrm{-}\bar{K^0}}
\newc{\solar}{\Delta m^2_{\odot}}
\newc{\atmos}{\Delta m^2_{atm}}
\newc{\mnu}{m_{\nu}}

\newc{\lam}{\lambda}

\newc{\mM}{\mathcal{M}}
\newc{\lag}{\mathcal{L}}

\begin{document}

\preprint{CAVENDISH-HEP-2009-03}
\preprint{DAMTP-2009-15}
\preprint{DO-TH-09/01}

\title{Large Hadron Collider probe of supersymmetric neutrinoless double beta
  decay mechanism} 

\author{B.~C.~Allanach}
\email[]{b.c.allanach@damtp.cam.ac.uk}
\affiliation{DAMTP, University of Cambridge, Wilberforce Road, Cambridge, CB3
  0WA, United Kingdom}

\author{C.~H.~Kom}
\email[]{kom@hep.phy.cam.ac.uk}
\affiliation{Cavendish Laboratory, J.J. Thomson Avenue, Cambridge CB3 0HE,
  United Kingdom}

\author{H.~P\"as}
\email[]{heinrich.paes@uni-dortmund.de}
\affiliation{Fakult\"at f\"ur Physik, Technische Universit\"at Dortmund, D-44221, Dortmund, Germany}

\date{\today}

\begin{abstract}
In the minimal supersymmetric extension to the Standard Model, a non-zero
lepton number violating coupling $\lam'_{111}$ predicts both neutrinoless
double  beta decay and resonant single slepton production at the LHC\@. We
show that, 
in this case, if neutrinoless double beta decay is discovered in the next
generation of experiments, there exist good prospects to observe single
slepton production at the LHC\@. Neutrinoless double beta decay could otherwise
result from a different source (such as a non-zero Majorana neutrino mass). 
Resonant single slepton production 
at the LHC can therefore discriminate between the 
$\lam'_{111}$ neutrinoless double beta decay mechanism and others.
\end{abstract}

\pacs{12.60.Jy, 13.15.tg, 14.80.Ly}

\maketitle


Neutrinoless double beta decay ($\betadecay$) corresponds to an atomic nucleus
changing two of its neutrons into 
protons, while emitting two electrons. 
At the quark level, the $\betadecay$ process corresponds to the simultaneous
transition of two down quarks (in different neutrons)
into two up-quarks and two electrons, but {\em without}\/ associated production
of any neutrinos. Thus, the $\betadecay$ process is lepton number violating
(LNV). 
$\betadecay$ has so far not been observed;
the most stringent lower limit on the $\Ge$ $\betadecay$ half life was
measured in the Heidelberg-Moscow experiment
\cite{Baudis:1999xd,Kleingrothaus:2000sn} to be 
\eqa \label{eq:halflifelimit}
\bhalflife &\geq& 1.9\cdot 10^{25} \textrm{yrs}.
\qea
Coverage by a couple of additional orders of magnitude is expected by planned
experiments in the coming years~\cite{Avignone:2007fu,Aalseth:2004hb}. The Standard Model conserves lepton number and
so predicts a zero rate for this process. A discovery of a non-zero rate
would then prompt the question: what beyond the Standard Model physics is
responsible for it? In this letter, we 
discuss two leading possibilities, Majorana neutrino
masses and supersymmetric particle exchange, pointing out how data from the
Large Hadron Collider (LHC) can favor or disfavor the latter possibility. 

\begin{figure}
    \includegraphics[scale=1]{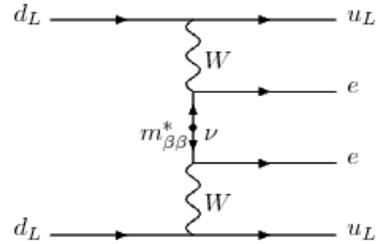}
\caption{Majorana neutrino mass induced neutrinoless double beta
  decay hard sub-process. \label{fig:mee0vbb}}
\end{figure}

The experimental 
observations of neutrino oscillations has
lead to the realization that at least two of the three known neutrinos have
masses~\cite{GonzalezGarcia:2007ib}. Thus, the
Standard Model, which predicts zero neutrino mass, must be augmented in some
way to account for such masses.  
Neutrino masses may or may not induce $\betadecay$ depending on whether they
are Majorana or Dirac masses, respectively. A Lagrangian for a LNV Majorana
neutrino mass is 
\begin{equation}
{\mathcal L}_{M}=\frac{1}{2}m_{\beta \beta} \overline{\nu^c}\nu + h.c.,
\label{majMass}
\end{equation}
where $\nu$ is the neutrino originating from the left-handed first generation
lepton electroweak 
doublet, and the $c$ superscript denotes charge conjugation.  
A Feynman diagram for the induced $\betadecay$ is shown in
Fig.~\ref{fig:mee0vbb}. 
It leads to an inverse half-life of
\begin{equation}
\Big[\bhalflife\Big]^{-1}= G_{01}\left|\frac{\mee}{m_e}M_{\nu}\right|^2,
\label{halfmnu}
\end{equation}
where $G_{01}=7.93\phantom{0}10^{-15}\textrm{yr}^{-1}$
\cite{Pantis:1996py} is a precisely calculable phase space factor, $m_e$ is
the electron mass and
$M_{\nu}$ denotes the nuclear matrix element (NME) for the process in
Fig.~\ref{fig:mee0vbb}. We shall use $M_{\nu}=2.8$ \cite{Simkovic:1999re} for
$\Ge$, but it should be noted that the uncertainty 
in the theoretical prediction of such nuclear matrix elements could be as
large as a factor of 3.
Eq.~\ref{halfmnu} then implies that, assuming $\betadecay$ is due solely to a
Majorana neutrino mass, 
\eqa \label{eq:scaledmeebound}
\frac{\mee}{460\textrm{meV}} &=& \Big(\frac{1.9\cdot 10^{25}\textrm{yr}}{\bhalflife({\rm min})}\Big)^{1/2}.
\qea

Several other possibilities of LNV processes that induce $\betadecay$ have been
discussed in the literature, including one attractive alternative where it is
mediated by the exchange of sparticles in supersymmetric models
with R-parity violation
\cite{Hirsch:1995ek,Faessler:1998qv,Hirsch:1995cg,Pas:1998nn,Faessler:2007nz}.
We shall focus on this possibility  in this letter.


The superpotential term
\begin{equation}
W = \lam'_{111} \hat L \hat Q \hat D^c \label{lamp}
\end{equation}
may induce $\betadecay$ and is allowed by the gauge symmetries of the minimal
supersymmetric standard model. $\hat Q$, $\hat L_i$ and $\hat D^c$ denote the
superfield containing left-handed quark doublet, left-handed
lepton doublet 
and charge conjugated right-handed down quark fields respectively (all
being of the first generation). Typically, one imposes a discrete symmetry on
the model in order to maintain proton stability. Such a symmetry may allow for
the 
presence of the term in Eq.~\ref{lamp} (for example baryon triality)
or ban it, as in the case of
$R-$parity \cite{Ibanez:1991hv}. We shall consider the former possibility here. 

\begin{figure}
    \includegraphics[scale=1]{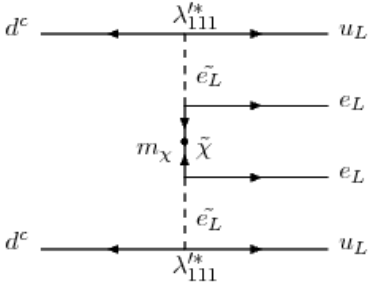}
\caption{Example Feynman diagram leading to neutrinoless double beta decay, 
mediated by supersymmetric particles. Several other such tree-level diagrams
are taken into account. \label{rpvnubb}}
\end{figure}

The interaction in Eq.~\ref{lamp} mediates
$\betadecay$ by processes such as the one shown in Fig.~\ref{rpvnubb}.
Following the notation of \cite{Hirsch:1995ek}, the effective Lagrangian with
$\lam'_{111}$ in the direct $R-$parity violating $\betadecay$ process
involving exchange of three supersymmetric (SUSY) particles is 
\eqa
\lag^{eff,\,\Delta L_e=2}_{\lam'_{111}\lam'_{111}}(x) = \frac{G^2_{F}}{2}m_p^{-1}[\bar{e}(1+\gamma_5)e^c] \nonumber \\
\times \Big[\eta (J_{PS}J_{PS}-\frac{1}{4}J^{\mu\nu}_T J_{T\mu\nu}) + \eta'J_{PS}J_{PS}\Big],
\qea
where 
\eqa
\eta=a\frac{\lam'^2_{111}}{G^2_F}\frac{m_P}{\Lambda^5_{SUSY}}, \
\eta'=b\frac{\lam'^2_{111}}{G^2_F}\frac{m_P}{\Lambda^5_{SUSY}}.
\qea
In the above expressions, $J_{PS}$ and $J^{\mu\nu}_T$ are the pseudo-scalar and tensor 
quark currents respectively.  The coefficients $a,b$ include factors 
coming from gauge couplings and mass matrix rotations, and $\Lambda_{SUSY}$
is the approximate mass scale of the sparticles being exchanged.  

The inverse half-life generated by $\lam'_{111}$ is
\begin{equation} \label{eq:halflifeformula}
\Big[\bhalflife\Big]^{-1} = G_{01}\left| M_{\lam'_{111}} \right|^2,
\end{equation}
where $M_{\lam'_{111}}$ denotes the relevant matrix
element, obtained from
Refs.~\cite{Hirsch:1995ek,Faessler:1998qv,Hirsch:1995cg,Pas:1998nn,Faessler:2007nz}, and is given by
\eqa
M_{\lam'_{111}} &=& \eta M^{2N}_{\tilde{g}} + \eta' M^{2N}_{\tilde{f}} \nonumber \\
&&+ \Big(\eta + \frac{5}{8}\eta'\Big) (\frac{4}{3}M^{1\pi} + M^{2\pi}),
\qea
with $M^{2N}_{\tilde{g},\tilde{f}}$ and $M^{1\pi,2\pi}$ denote NME
contributions from 2 nucleon lepton decay and pion exchange modes respectively.  
The numerical values of the NMEs we shall use are displayed in table~\ref{tab:NMEinput}.
We refer interested readers to \cite{AKP2009} for a more detailed discussion.

\begin{table}[t]
  \centering
  \begin{tabular}{|cccc|}
    \hline
    $M^{2N}_{\tilde{g}}$   &$M^{2N}_{\tilde{f}}$   &$M^{1\pi}$   &$M^{2\pi}$ \\
    \hline
    283 \cite{Hirsch:1995ek} &13.2 \cite{Hirsch:1995ek} &-18.2 \cite{Faessler:1998qv} &-601 \cite{Faessler:1998qv}\\
    \hline
  \end{tabular}
  \caption[]{Nuclear matrix elements of $\Ge$ used.  For model details of the NME calculations, we refer readers to the literature.}\label{tab:NMEinput}
\end{table}

The experimental lower bound in Eq.~\ref{eq:halflifelimit} then leads to the
approximate limit 
\cite{Hirsch:1995ek,Faessler:1998qv} 
\begin{equation} 
|\lam'_{111}| \lesssim 5 \cdot 10^{-4}\Big(\frac{\Lambda_{SUSY}}{100\textrm{GeV}}\Big)^{2.5}.\label{eq:111directdecaylimit}
\end{equation}

Couplings such as Eq.~\ref{lamp} also lead
to  loop-level Majorana left-handed neutrino masses~\cite{Hall:1983id}
\eqa\label{eq:mee}
\mee &\simeq & \frac{3m_d}{8\pi^2}\frac{\lambda'^2_{111}m^2_{\tilde{d}_{LR}}}{m^2_{\tilde{d}_{LL}}-m^2_{\tilde{d}_{RR}}}\textrm{ln}\Big(\frac{m^2_{\tilde{d}_{LL}}}{m^2_{\tilde{d}_{RR}}}\Big),
\qea
where $m_d$ is the down quark mass, while $m^2_{\tilde{d}_{LL,LR,RR}}$ are
entries in the first generation down squark mass squared matrix. 
Thus there is potentially an additional contribution to $\betadecay$ from the
induced neutrino mass in Eq.~\ref{eq:mee}. 
Through the
parameter space that we consider $|M_{\lam'_{111}}|/|M_{\mee}| > 20$,
where $M_{\mee}\equiv \mee M_\nu / m_e$, and so we
may neglect the contribution coming from induced Majorana 
neutrino masses. 

We pick an
illustrative scheme of supersymmetry breaking: the so-called mSUGRA
assumption. 
The following set
of parameters is defined: 
$\MO=[40,1000]$ GeV, $\Mhalf=[40,1000]$ GeV, $A_0=0$ $\tanb=10$, $\sgnmu=+1$,
where $\MO$, $\Mhalf$ and $A_{0}$ are the universal scalar, gaugino, and
trilinear soft SUSY breaking parameters defined at the electroweak gauge
coupling unification scale $M_{X}\sim 2.0\cdot 10^{16}\textrm{GeV}$, $\tanb$
is the ratio of the Higgs vacuum expectation values $v_u/v_d$, and $\sgnmu$ is
the sign of the bilinear Higgs parameter in the superpotential. 

For large enough $\lam'_{111}$, resonant production of 
a single slepton of the first generation\footnote{We will refer to this process
  simply as `single slepton production' unless specified otherwise.} may be
observed at the LHC.    
Neglecting finite width effects, the color and spin-averaged parton total
cross section of a single slepton production is~\cite{Dimopoulos:1988jw}
\eqa
\hat{\sigma} &=& \frac{\pi}{12\hat{s}}|\lam'_{111}|^2\delta(1-\frac{m^2_{\tilde{l}}}{\hat{s}}),
\qea
where $\hat{s}$ is the partonic center of mass energy, and
$m_{\tilde{l}}$ is the mass of the resonant slepton. 
Including effects from
parton distribution functions, we find that the total cross section for
$\sigma (pp \rightarrow \tilde l) \propto |\lam'_{111}|^2 / m_{\tilde l}^3 $
to a good approximation
in the parameter region of interest.

At low slepton masses, the stringent bound in
Eq.~\ref{eq:111directdecaylimit}
from $\betadecay$ renders such a process unobservable at the LHC.
 We believe that this has precluded any study of single slepton 
 production of the \emph{first} generation at the LHC via $\lam'_{111}$.
However, from eq.~(\ref{eq:111directdecaylimit}), we see that, applying the
bound on $\lam'_{111}$ coming from non-observation of $\betadecay$, 
$\sigma< c \Lambda_{SUSY}^2$ where $c$ is a constant, and so at
higher values of the supersymmetric masses, larger cross-sections may be
allowed due to a much larger allowable $\lam'_{111}$. It is this possibility
that we exploit here. 

A closely related process, LHC second generation slepton production, followed
by decay into  
like-sign di-muon pairs, was studied in Ref.~\cite{Dreiner:2000vf}.
Such a process is predicted by the superpotential term $\lam'_{211} \hat L_2
\hat Q \hat D^c$, where $L_2$ is a chiral superfield containing the second
generation 
left-handed lepton doublet. $\lam'_{211}$ does not predict $\betadecay$ and so
it may take a somewhat larger value than $\lam'_{111}$ for a given set of
supersymmetric particle masses. LHC detectors do not have wildly differing
acceptances and efficiencies for electrons as compared with muons, and so 
we use the results of Ref.~\cite{Dreiner:2000vf} (which does not include
detector effects anyway)
as an estimate for the
search reach for first generation single slepton production, followed by decay
into like-sign electrons, by simply making the replacements $\lam'_{211}
\rightarrow \lam'_{111}$ and $\mu \rightarrow e$. 
\begin{figure}
\includegraphics[scale=1]{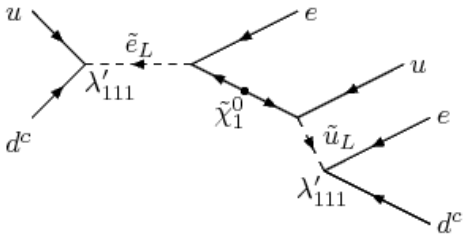}
\caption{Example of single selectron production at the LHC, followed by
  subsequent cascade decay. \label{fig:singsel}}
\end{figure}
A Feynman diagram leading to our signal (like-sign di-electron pairs and two hard
jets, with no missing energy) is shown in Fig.~\ref{fig:singsel}.

Like Ref.~\cite{Dreiner:2000vf}, we 
assume 10$\,\ifb$ of LHC integrated luminosity
at a centre of mass energy of 14 TeV. 
\begin{figure}[t!]
\begin{picture}(200,180)(0,0)
\put(-10,0){\includegraphics[width=0.42\textwidth]{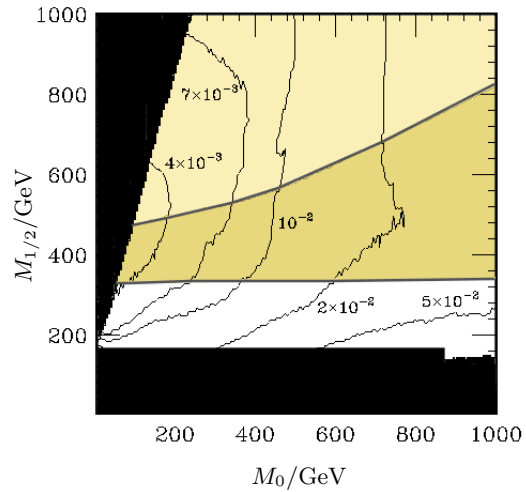}}
    \put(80,4){$\MO$/GeV}
    \put(-10,80){\rotatebox{90}{$\Mhalf$/GeV}}
\end{picture}
  \caption{mSUGRA parameter space in which
  single slepton production may be observed at the LHC for $\tan \beta=10$, 
  $A_0=0$ and 10$\ifb$ of integrated luminosity. 
  In the top left-hand black triangle, the stau is the LSP, a case not covered
  by this analysis. The bottom black region is ruled out by direct
  search constraints. The labelled contours are extracted from
  Ref.~\cite{Dreiner:2000vf}, and show the search reach given by the labelled
  value of $\lam'_{111}$. The white, dark-shaded and light-shaded regions
  show that observation of single slepton production at the $5\sigma$ level
  would imply     $\bhalflife < 1.9\cdot 10^{25} \textrm{yrs}$,  $100> \bhalflife/10^{25}  \textrm{yrs} > 1.9$ and
    $\bhalflife > 1 \times 10^{27} \textrm{yrs}$, respectively.
  \label{fig:lamp111_Disc}}
\end{figure}
Fig.~\ref{fig:lamp111_Disc} shows regions of the
$\MO-\Mhalf$ plane where single slepton production may be observed via
like-sign electrons plus two jets, including backgrounds from both the
Standard Model and from sparticle pair production. The cuts are as in
Ref.~\cite{Dreiner:2000vf}. 
In the white region, single
slepton production by 
$\lam'_{111}$ could not be observed without violating the current bound upon
$\bhalflife$. 
The darker shaded region shows where the observation of single
slepton production at $5\sigma$ above background implies that $\betadecay$ is within the reach of the
next generation of experiments, which should be able to probe $\bhalflife < 1
\times 10^{27}\textrm{yrs}$~\cite{Avignone:2007fu,Aalseth:2004hb}. Conversely, if $\betadecay$ is discovered by the next
generation of experiments, we should expect single slepton production to be
observable and test the $\lam'_{111}$ hypothesis. We do not expect $A_0$ or
$\tan \beta$ to
affect the shape of the 
regions much, since they have a negligible effect on the
selectron mass and the couplings in the relevant Feynman diagrams. 
In the light shaded (upper) region, a 5$\sigma$ single slepton discovery at
the LHC implies that
the next generation of experiments would not be able to observe $\betadecay$.
Conversely, if $\betadecay$ is within reach of the next generation of
experiments, the LHC would see single slepton production signal in this region
at greater than $5 \sigma$ significance.

\begin{figure}[t!]
\begin{picture}(200,180)(0,0)
\put(-45,0){\includegraphics[width=0.55\textwidth]{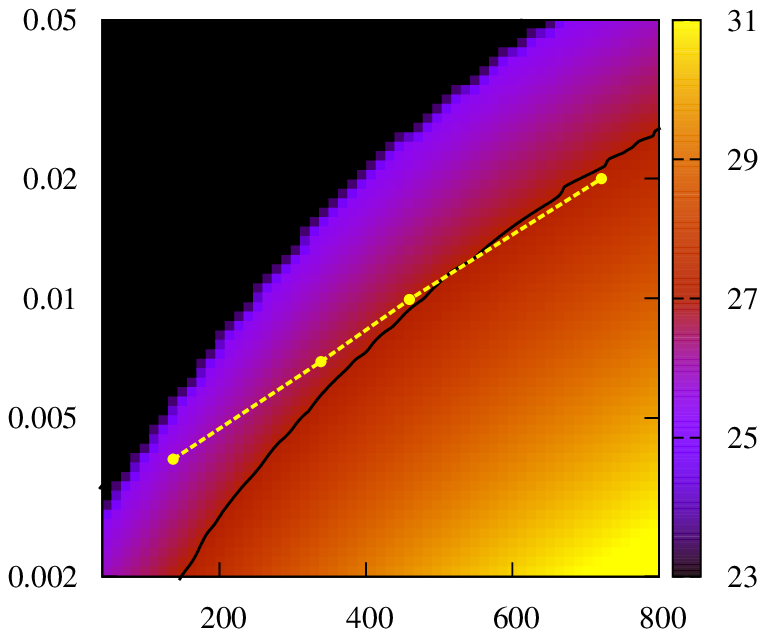}}
    \put(80,15){$\MO$/GeV}
    \put(-5,80){\rotatebox{90}{$\lam'_{111}$}}
    \put(145,170){$\textrm{log}_{10}(\bhalflife)$}
\end{picture}
  \caption{Comparison of $\bhalflife$ and single slepton discovery reach as a
    function of $\lam'_{111}$ along the mSUGRA slope $M_{1/2}=300\mbox{~GeV}+0.6M_{0}$,
    with $\AO=0$, $\tanb=10$ and $\textrm{sgn}\mu=1$.  The black region on the
    top left corner is ruled out by $\betadecay$.  The region above the solid
    black line is accessible in near future $\betadecay$ experiments, whereas
    the light dotted line shows the lower limit of $\lam'_{111}$ for single
    slepton production to be discoverable at the
    LHC.  \label{fig:lamp111vsM0}} 
\end{figure}
We show in Fig.~\ref{fig:lamp111vsM0} the variation of the discovery reach of $\lam'_{111}$
with $M_{0}$ along the line $M_{1/2}=300 \mbox{~GeV} + 0.6M_{0}$ in
Fig.~\ref{fig:lamp111_Disc}.  Above the dotted light line, single slepton
production  
will be observed at the LHC\@. We see from the figure that for nearly all of the
parameter space where $\betadecay$ can be measured by the next generation of
experiments,  the LHC would provide a confirmation of the supersymmetric
origin of the signal by observing single slepton production at the 5$\sigma$
level. 

In summary, we have discussed the interplay between 
neutrinoless double beta decay and single slepton
production at the LHC in $R-$parity violating supersymmetry. 
Should neutrinoless double beta decay be observed in the next round of
experiments, one would like to interpret which physics would lead to 
the observation. We have considered the exchange of supersymmetric particles via
the lepton number violating interaction in Eq.~\ref{lamp}.
The observation of single slepton production could discriminate between this
possibility and others (for example a Majorana neutrino mass).
Fig.~\ref{fig:lamp111_Disc} shows that much of the
parameter space allowed by $\betadecay$ in simple models of 
supersymmetry breaking predicts observable single slepton production at the
LHC\@. It also shows that if
the next round of experiments observe the $\betadecay$ process, the LHC has a
very good chance of observing single slepton production with only 10
fb$^{-1}$ of integrated luminosity, assuming that $\betadecay$ is induced by a
$\lam'_{111}$ coupling. Conversely, non-observation 
of single slepton production could then discriminate against 
the $\lam'_{111}$ mechanism.
In general, one may enquire whether {\em both}\/ Majorana neutrino masses and
$\lam'_{111}$ contribute simultaneously and non-negligibly to $\betadecay$. 
Detailed LHC measurements of the kinematics in single slepton production
would constrain  
the mSUGRA parameters, and the total cross-section could then give information
about the size of $|\lam'_{111}|$. The LHC information could be combined to
predict an associated 
inverse $\bhalflife$ coming from $\lam'_{111}$, which could
be compared with the experimental measurement of $\bhalflife$ in order to see
if additional contributions were necessary. It will be interesting in future
studies to see how accurate such an inference could be, assuming matrix
element uncertainties can be kept under control.

\begin{acknowledgments}
This work has been partially supported by STFC\@.  
BCA received partial funding from the Gambrinus
Fellowship, 
CHK from the
Hutchison Whampoa Dorothy 
Hodgkin Postgraduate Award 
and HP by the EU project ILIAS N6 WP1.
We thank the members of the Cambridge SUSY working group,
G.~Hiller, M.~Hirsch and R.~Mohapatra for 
valuable conversations.  BCA and CHK also thank the Technische Universit\"at 
Dortmund, and HP thanks the University of Cambridge for hospitality offered
 while part of this work was carried out. 
\end{acknowledgments}



\begin{thebibliography}{17}
\expandafter\ifx\csname natexlab\endcsname\relax\def\natexlab#1{#1}\fi
\expandafter\ifx\csname bibnamefont\endcsname\relax
  \def\bibnamefont#1{#1}\fi
\expandafter\ifx\csname bibfnamefont\endcsname\relax
  \def\bibfnamefont#1{#1}\fi
\expandafter\ifx\csname citenamefont\endcsname\relax
  \def\citenamefont#1{#1}\fi
\expandafter\ifx\csname url\endcsname\relax
  \def\url#1{\texttt{#1}}\fi
\expandafter\ifx\csname urlprefix\endcsname\relax\def\urlprefix{URL }\fi
\providecommand{\bibinfo}[2]{#2}
\providecommand{\eprint}[2][]{\url{#2}}

\bibitem[{\citenamefont{Baudis et~al.}(1999)}]{Baudis:1999xd}
\bibinfo{author}{\bibfnamefont{L.}~\bibnamefont{Baudis}} \bibnamefont{et~al.},
  \bibinfo{journal}{Phys. Rev. Lett.} \textbf{\bibinfo{volume}{83}},
  \bibinfo{pages}{41} (\bibinfo{year}{1999}), \eprint{hep-ex/9902014}.

\bibitem[{\citenamefont{Klapdor-Kleingrothaus
  et~al.}(2001)}]{Kleingrothaus:2000sn}
\bibinfo{author}{\bibfnamefont{H.~V.} \bibnamefont{Klapdor-Kleingrothaus}}
  \bibnamefont{et~al.}, \bibinfo{journal}{Eur. Phys. J.}
  \textbf{\bibinfo{volume}{A12}}, \bibinfo{pages}{147} (\bibinfo{year}{2001}),
  \eprint{hep-ph/0103062}.

\bibitem[{\citenamefont{Avignone et~al.}(2008)\citenamefont{Avignone, Elliott,
  and Engel}}]{Avignone:2007fu}
\bibinfo{author}{\bibfnamefont{I.}~\bibnamefont{Avignone},
  \bibfnamefont{Frank~T.}}, \bibinfo{author}{\bibfnamefont{S.~R.}
  \bibnamefont{Elliott}}, \bibnamefont{and}
  \bibinfo{author}{\bibfnamefont{J.}~\bibnamefont{Engel}},
  \bibinfo{journal}{Rev. Mod. Phys.} \textbf{\bibinfo{volume}{80}},
  \bibinfo{pages}{481} (\bibinfo{year}{2008}), \eprint{0708.1033}.

\bibitem[{\citenamefont{Aalseth et~al.}(2004)}]{Aalseth:2004hb}
\bibinfo{author}{\bibfnamefont{C.}~\bibnamefont{Aalseth}} \bibnamefont{et~al.}
  (\bibinfo{year}{2004}), \eprint{hep-ph/0412300}.

\bibitem[{\citenamefont{Gonzalez-Garcia and
  Maltoni}(2008)}]{GonzalezGarcia:2007ib}
\bibinfo{author}{\bibfnamefont{M.~C.} \bibnamefont{Gonzalez-Garcia}}
  \bibnamefont{and} \bibinfo{author}{\bibfnamefont{M.}~\bibnamefont{Maltoni}},
  \bibinfo{journal}{Phys. Rept.} \textbf{\bibinfo{volume}{460}},
  \bibinfo{pages}{1} (\bibinfo{year}{2008}), \eprint{0704.1800}.

\bibitem[{\citenamefont{Pantis et~al.}(1996)\citenamefont{Pantis, Simkovic,
  Vergados, and Faessler}}]{Pantis:1996py}
\bibinfo{author}{\bibfnamefont{G.}~\bibnamefont{Pantis}},
  \bibinfo{author}{\bibfnamefont{F.}~\bibnamefont{Simkovic}},
  \bibinfo{author}{\bibfnamefont{J.~D.} \bibnamefont{Vergados}},
  \bibnamefont{and} \bibinfo{author}{\bibfnamefont{A.}~\bibnamefont{Faessler}},
  \bibinfo{journal}{Phys. Rev.} \textbf{\bibinfo{volume}{C53}},
  \bibinfo{pages}{695} (\bibinfo{year}{1996}), \eprint{nucl-th/9612036}.

\bibitem[{\citenamefont{Simkovic et~al.}(1999)\citenamefont{Simkovic, Pantis,
  Vergados, and Faessler}}]{Simkovic:1999re}
\bibinfo{author}{\bibfnamefont{F.}~\bibnamefont{Simkovic}},
  \bibinfo{author}{\bibfnamefont{G.}~\bibnamefont{Pantis}},
  \bibinfo{author}{\bibfnamefont{J.~D.} \bibnamefont{Vergados}},
  \bibnamefont{and} \bibinfo{author}{\bibfnamefont{A.}~\bibnamefont{Faessler}},
  \bibinfo{journal}{Phys. Rev.} \textbf{\bibinfo{volume}{C60}},
  \bibinfo{pages}{055502} (\bibinfo{year}{1999}), \eprint{hep-ph/9905509}.

\bibitem[{\citenamefont{Hirsch et~al.}(1996{\natexlab{a}})\citenamefont{Hirsch,
  Klapdor-Kleingrothaus, and Kovalenko}}]{Hirsch:1995ek}
\bibinfo{author}{\bibfnamefont{M.}~\bibnamefont{Hirsch}},
  \bibinfo{author}{\bibfnamefont{H.~V.} \bibnamefont{Klapdor-Kleingrothaus}},
  \bibnamefont{and} \bibinfo{author}{\bibfnamefont{S.~G.}
  \bibnamefont{Kovalenko}}, \bibinfo{journal}{Phys. Rev.}
  \textbf{\bibinfo{volume}{D53}}, \bibinfo{pages}{1329}
  (\bibinfo{year}{1996}{\natexlab{a}}), \eprint{hep-ph/9502385}.

\bibitem[{\citenamefont{Faessler et~al.}(1998)\citenamefont{Faessler,
  Kovalenko, and Simkovic}}]{Faessler:1998qv}
\bibinfo{author}{\bibfnamefont{A.}~\bibnamefont{Faessler}},
  \bibinfo{author}{\bibfnamefont{S.}~\bibnamefont{Kovalenko}},
  \bibnamefont{and} \bibinfo{author}{\bibfnamefont{F.}~\bibnamefont{Simkovic}},
  \bibinfo{journal}{Phys. Rev.} \textbf{\bibinfo{volume}{D58}},
  \bibinfo{pages}{115004} (\bibinfo{year}{1998}), \eprint{hep-ph/9803253}.

\bibitem[{\citenamefont{Hirsch et~al.}(1996{\natexlab{b}})\citenamefont{Hirsch,
  Klapdor-Kleingrothaus, and Kovalenko}}]{Hirsch:1995cg}
\bibinfo{author}{\bibfnamefont{M.}~\bibnamefont{Hirsch}},
  \bibinfo{author}{\bibfnamefont{H.~V.} \bibnamefont{Klapdor-Kleingrothaus}},
  \bibnamefont{and} \bibinfo{author}{\bibfnamefont{S.~G.}
  \bibnamefont{Kovalenko}}, \bibinfo{journal}{Phys. Lett.}
  \textbf{\bibinfo{volume}{B372}}, \bibinfo{pages}{181}
  (\bibinfo{year}{1996}{\natexlab{b}}), \eprint{hep-ph/9512237}.

\bibitem[{\citenamefont{Pas et~al.}(1999)\citenamefont{Pas, Hirsch, and
  Klapdor-Kleingrothaus}}]{Pas:1998nn}
\bibinfo{author}{\bibfnamefont{H.}~\bibnamefont{P\"as}},
  \bibinfo{author}{\bibfnamefont{M.}~\bibnamefont{Hirsch}}, \bibnamefont{and}
  \bibinfo{author}{\bibfnamefont{H.~V.} \bibnamefont{Klapdor-Kleingrothaus}},
  \bibinfo{journal}{Phys. Lett.} \textbf{\bibinfo{volume}{B459}},
  \bibinfo{pages}{450} (\bibinfo{year}{1999}), \eprint{hep-ph/9810382}.

\bibitem[{\citenamefont{Faessler et~al.}(2008)\citenamefont{Faessler, Gutsche,
  Kovalenko, and Simkovic}}]{Faessler:2007nz}
\bibinfo{author}{\bibfnamefont{A.}~\bibnamefont{Faessler}},
  \bibinfo{author}{\bibfnamefont{T.}~\bibnamefont{Gutsche}},
  \bibinfo{author}{\bibfnamefont{S.}~\bibnamefont{Kovalenko}},
  \bibnamefont{and} \bibinfo{author}{\bibfnamefont{F.}~\bibnamefont{Simkovic}},
  \bibinfo{journal}{Phys. Rev.} \textbf{\bibinfo{volume}{D77}},
  \bibinfo{pages}{113012} (\bibinfo{year}{2008}), \eprint{0710.3199}.

\bibitem[{\citenamefont{Ibanez and Ross}(1991)}]{Ibanez:1991hv}
\bibinfo{author}{\bibfnamefont{L.~E.} \bibnamefont{Ibanez}} \bibnamefont{and}
  \bibinfo{author}{\bibfnamefont{G.~G.} \bibnamefont{Ross}},
  \bibinfo{journal}{Phys. Lett.} \textbf{\bibinfo{volume}{B260}},
  \bibinfo{pages}{291} (\bibinfo{year}{1991}).

\bibitem[{\citenamefont{Allanach et~al.}(2009)\citenamefont{Allanach, Kom, and
  Pas}}]{AKP2009}
\bibinfo{author}{\bibfnamefont{B.~C.} \bibnamefont{Allanach}},
  \bibinfo{author}{\bibfnamefont{C.-H.} \bibnamefont{Kom}}, \bibnamefont{and}
  \bibinfo{author}{\bibfnamefont{H.}~\bibnamefont{P\"as}} (\bibinfo{year}{2009}),
  \eprint{to appear}.

\bibitem[{\citenamefont{Hall and Suzuki}(1984)}]{Hall:1983id}
\bibinfo{author}{\bibfnamefont{L.~J.} \bibnamefont{Hall}} \bibnamefont{and}
  \bibinfo{author}{\bibfnamefont{M.}~\bibnamefont{Suzuki}},
  \bibinfo{journal}{Nucl. Phys.} \textbf{\bibinfo{volume}{B231}},
  \bibinfo{pages}{419} (\bibinfo{year}{1984}).

\bibitem[{\citenamefont{Dimopoulos and Hall}(1988)}]{Dimopoulos:1988jw}
\bibinfo{author}{\bibfnamefont{S.}~\bibnamefont{Dimopoulos}} \bibnamefont{and}
  \bibinfo{author}{\bibfnamefont{L.~J.} \bibnamefont{Hall}},
  \bibinfo{journal}{Phys. Lett.} \textbf{\bibinfo{volume}{B207}},
  \bibinfo{pages}{210} (\bibinfo{year}{1988}).

\bibitem[{\citenamefont{Dreiner et~al.}(2001)\citenamefont{Dreiner, Richardson,
  and Seymour}}]{Dreiner:2000vf}
\bibinfo{author}{\bibfnamefont{H.~K.} \bibnamefont{Dreiner}},
  \bibinfo{author}{\bibfnamefont{P.}~\bibnamefont{Richardson}},
  \bibnamefont{and} \bibinfo{author}{\bibfnamefont{M.~H.}
  \bibnamefont{Seymour}}, \bibinfo{journal}{Phys. Rev.}
  \textbf{\bibinfo{volume}{D63}}, \bibinfo{pages}{055008}
  (\bibinfo{year}{2001}), \eprint{hep-ph/0007228}.

\end{thebibliography}
\end{document}